\documentstyle[pra,aps,amsfonts,multicol]{revtex}
\newcommand{\identity}{\openone}
\newcommand{\C}{{\Bbb C}}
\newcommand{\trace}{\mathop{\rm Tr}\nolimits}
\newcommand{\bra}[1]{\langle#1|}
\newcommand{\ket}[1]{|#1\rangle}
\newcommand{\qed}{\hfill$\Box$}
\newcommand{\diag}{\mathop{\rm diag}\nolimits}

\newtheorem{theorem}{Theorem}
\newtheorem{lemma}{Lemma}
\draft
\title{Maximally entangled mixed states of two qubits}
\author{Frank Verstraete, Koenraad Audenaert, Tijl De Bie, Bart De Moor}
\address{Katholieke Universiteit Leuven,
Department of Electrical Engineering, Research Group SISTA\\
Kard. Mercierlaan 94, B-3001 Leuven, Belgium }
\begin{document}

\pagestyle{plain} \pagenumbering{arabic}

\maketitle
\begin{abstract}
We consider mixed states of two qubits and show under which
global unitary operations their entanglement is maximized. This
leads to a class of states that is a generalization of the Bell
states. Three measures of entanglement are considered:
entanglement of formation, negativity and relative entropy of
entanglement. Surprisingly all states that maximize one measure
also maximize the others. We will give a complete characterization
of these generalized Bell states and prove that these states for
fixed eigenvalues are all equivalent under local unitary
transformations. We will furthermore characterize all nearly
entangled states closest to the maximally mixed state and derive
a new lower bound on the volume of separable mixed states.
\end{abstract}
\pacs{03.65.Bz, 03.67.-a, 89.70.+c}

\begin{multicols}{2}[]
\narrowtext

In this paper we investigate how much entanglement in a mixed two
qubit system can be created by global unitary transformations. The
class of states for which no more entanglement can be created by
global unitary operations is clearly a generalization of the
class of Bell states to mixed states, and gives strict bounds on
how the mixedness of a state limits its entanglement. This
question is of considerable interest as entanglement is the magic
ingredient of quantum information theory and experiments always
deal with mixed states. Recently, Ishizaka and Hiroshima
\cite{ishizaka} independently considered the same question. They
proposed a class of states and conjectured that the entanglement
of formation \cite{wootters} and the negativity \cite{zyczkowski}
of these states could not be increased by any global unitary
operation. Here we rigorously prove their conjecture and
furthermore prove that the states they proposed are the only ones
having the property of maximal entanglement.

Closely related to the issue of generalized Bell states is the
question of characterizing the set of separable density matrices
\cite{braunstein}, as the entangled states closest to the
maximally mixed state necessarily have to belong to the proposed
class of maximal entangled mixed states.  We can thus give a
complete characterization of all nearly entangled states lying on
the boundary of the sphere of separable states surrounding the
maximally mixed state. As a byproduct this gives an alternative
derivation of the well known result of Zyczkowski et al.
\cite{zyczkowski} that all states for which the inequality ${\rm
Tr}(\rho^2)\leq 1/3$ holds are separable.

The original motivation of this paper was the following question:
given a single quantum mechanical system consisting of two
unentangled spin-$1/2$ systems, i.e.\ two qubits, one in a pure
state and another in a maximally mixed state, does there exist a
global unitary transformation on both qubits such that they become
entangled? Surprisingly, the answer is yes. In this paper we solve
the more general question: how can one maximize the entanglement
of an arbitrary mixerd state of two qubits using only unitary
operations. If not only unitary operations but also measurements
were allowed, it is clear that a Von Neumann measurement in the
Bell basis would immediatly yield a singlet. Here however we
restrict ourselves to unitary operations. Obviously, these unitary
operations must be global ones, that is, acting on the system as a
whole, since any reasonable measure of entanglement must be
invariant under local unitary operations, acting only on single
qubits. As measures of entanglement, the entanglement of formation
(EoF) \cite{wootters}, the negativity \cite{zyczkowski} and the
relative entropy of entanglement \cite{vedral} were chosen.

The entanglement of formation of mixed states is defined
variationally as $E_f(\rho)=\min_{\{\psi_i\}}\sum_ip_iE(\psi_i)$
where $\rho=\sum_i p_i\psi_i\psi_i^{\dagger}$. For $2\times 2$
systems the EoF is well-characterized by introducing the
concurrence $C$ \cite{wootters}:
\begin{eqnarray}E_f(\rho)&=&f(C(\rho))=H\left(\frac{1+\sqrt{1-C^2}}{2}\right)\\
C(\rho)&=&\max(0,\sigma_1-\sigma_2-\sigma_3-\sigma_4).\end{eqnarray}
Here $\{\sigma_i\}$ are the square roots of the eigenvalues of the
matrix $A$ arranged in decreasing order
\begin{eqnarray} A&=&\rho S \rho^* S\\
S&=&\sigma_y\otimes\sigma_y.\end{eqnarray} $H(x)$ is Shannon's
entropy function and $\sigma_y$ is the Pauli matrix. It can be
shown that $f(C)$ is convex and monotonously increasing. Using
some elementary linear algebra it is furthermore easy to prove
that the numbers $\{\sigma_i\}$ are equal to the singular values
\cite{horn85} of the matrix $\sqrt{\rho}^T S\sqrt{\rho}$. Here we
use the notation $\sqrt{\rho}=\Phi\Lambda^{1/2}$ given
$\Phi\Lambda\Phi^\dagger$, the eigenvalue decomposition of $\rho$.

The concept of negativity of a state is closely related to the
well-known Peres condition for separability of a state
\cite{peres}. If a state is separable (disentangled), then the
partial transpose of the state is again a valid state, i.e.\ it
is positive. For $2\times2$ systems, this condition is also
sufficient \cite{horodecki}. It turns out that the partial
transpose of a non-separable state has one negative eigenvalue.
From this, a measure for entanglement follows: the {\em
negativity} of a state \cite{zyczkowski} is equal to the trace
norm of its partial transpose. We will adopt the definition of
negativity as twice the absolute value of this negative
eigenvalue:
\begin{equation}E_N(\rho) = 2\max(0,-\lambda_4),\end{equation}
where $\lambda_4$ is the minimal eigenvalue of $\rho^{T_A}$. In
the case of two qubits, this is equivalent to the trace norm of
the partial transpose up to an affine mapping.

The relative entropy of entanglement was proposed by Vedral and
Plenio \cite{vedral} as a measure of entanglement motivated by the
classical concept of Kullback-Leibler distance between probability
distributions. This measure has very nice properties such as being
a good upper bound for the entanglement of distillation. It is
variationally defined as
\begin{equation}
E_R(\rho)=\min_{\sigma\in D}{\rm
Tr}\left(\rho\log\rho-\rho\log\sigma\right)
\end{equation}
where $D$ represents the convex set of all separable density
operators.

We now state our main result:
\begin{theorem}
Let the eigenvalue decomposition of $\rho$ be
\[\rho=\Phi\Lambda\Phi^\dagger\] where the eigenvalues
$\{\lambda_i\}$ are sorted in non-ascending order. The
entanglement of formation is maximized if and only if a global
unitary transformation of the form
\[U=\left(U_1\otimes
U_2\right)\left(\begin{array}{cccc}0&0&0&1\\1/\sqrt{2}&0&1/\sqrt{2}&0\\1/\sqrt{2}&0&-1/\sqrt{2}&0\\0&1&0&0\end{array}\right)D_\phi\Phi^{\dagger}\]
is applied to the system, where $U_1$ and $U_2$ are local unitary
operations and $D_\phi$ is a unitary diagonal matrix. This same
global unitary transformation is the unique transformation
maximizing the negativity and the relative entropy of
entanglement. The entanglement of formation and negativity of the
new state $\rho'=U\rho U^\dagger$ are then given by
\begin{eqnarray}E_f(\rho')&=&f\left(\max\left(0,\lambda_1-\lambda_3-2\sqrt{\lambda_2\lambda_4}\right)\right)\nonumber\\
E_N(\rho')&=&\max\left(0,\sqrt{(\lambda_1-\lambda_3)^2+(\lambda_2-\lambda_4)^2}-\lambda_2-\lambda_4\right)\nonumber\end{eqnarray}
respectively, while the expression for the relative entropy of
entanglement is given by
\begin{eqnarray*}
&&\hspace{-.7cm}E_R(\rho')={\rm
Tr}(\rho\log\rho)-\lambda_1\log((1-a)/2)-\\
&&\lambda_2\log((a+b+2(\lambda_2-\lambda_4))/4)-\\
&&\lambda_3\log((1-b)/2)-\lambda_4\log((a+b-2(\lambda_2-\lambda_4))/4)\\
a&=&(d-\sqrt{d^2-4(1-\lambda_1)(1-\lambda_3)(\lambda_2-\lambda_4)^2})/(2(1-\lambda_3))\\
b&=&(d-\sqrt{d^2-4(1-\lambda_1)(1-\lambda_3)(\lambda_2-\lambda_4)^2})/(2(1-\lambda_1))\\
d&=&\lambda_2+\lambda_4+(\lambda_2-\lambda_4)^2
\end{eqnarray*}
\end{theorem}
The class of generalized Bell states is defined as the states
$\rho'$ thus obtained. These states are the maximally entangled
mixed states (MEMS-states).

We now present the complete proof of this Theorem. The cases of
entanglement of formation, negativity and relative entropy of
entanglement will be treated independently. We start with the
entanglement of formation.

As the function $f(x)$ is monotonously increasing, maximizing the
EoF is equivalent to maximizing the concurrence. The problem is
now reduced to finding: \begin{equation} C_{\max} = \max_{U\in
U(4)} (0,\sigma_1-\sigma_2-\sigma_3-\sigma_4)\end{equation} with
$\{\sigma_i\}$ the singular values of \begin{equation} Q =
\Lambda^{1/2}\Phi^T U^T S U\Phi\Lambda^{1/2}.\end{equation} Now,
$\Phi$, $U$ and $S$ are unitary, and so is any product of them.
It then follows that \begin{equation}C_{max} \le \max_{V\in U(4)}
(0,\sigma_1-\sigma_2-\sigma_3-\sigma_4)\end{equation} with
$\{\sigma_i\}$ the singular values of $\Lambda^{1/2} V
\Lambda^{1/2}$. The inequality becomes an equality if there is a
unitary matrix $U$ such that the optimal $V$ can be written as
$\Phi^T U^T S U\Phi$. A necessary and sufficient condition for
this is that the optimal $V$ be symmetric ($V=V^T$): as $S$ is
symmetric and unitary, it can be written as a product $S_1^T
S_1$, with $S_1$ again unitary. This is known as the Takagi
factorization of $S$ \cite{horn85}. This factorization is not
unique: left-multiplying $S_1$ with a complex orthogonal matrix
$O$ ($O^T O=\identity$) also yields a valid Takagi factor. An
explicit form of $S_1$ is given by: \begin{equation}
S_1=\frac{1}{\sqrt{2}}\left(\begin{array}{cccc}0 & 1 & 1 & 0\\-1 & 0 & 0 & 1\\
0 & -i & i & 0\\i & 0 & 0 &i\end{array}\right).\end{equation} If
$V$ is symmetric it can also be factorized like this: $V=V_1^T
V_1$. It is now easy to see that any $U$ of the form
\begin{equation} U=S_1^\dagger O V_1 \Phi^\dagger,\end{equation} with $O$ real
orthogonal, indeed yields $V=V_1^TV_1$.

To proceed, we need two inequalities concerning singular values
of matrix products. Henceforth, singular values, as well as
eigenvalues will be sorted in non-ascending order. The following
inequality for singular values is well-known \cite{horn91}:
\begin{lemma}
Let $A\in M_{n,r}(\C)$, $B\in M_{r,m}(\C)$. Then, \begin{equation}
\sum_{i=1}^k \sigma_i(AB) \le \sum_{i=1}^k \sigma_i(A)
\sigma_i(B),\end{equation} for $k=1,\ldots,q=\min\{n,r,m\}$.
\end{lemma}
Less known is the following result by Wang and Xi \cite{wang97}:
\begin{lemma}
Let $A\in M_n(\C)$, $B\in M_{n,m}(\C)$, and $1\le i_1
<\cdots<i_k\le n$. Then \begin{equation}\sum_{t=1}^k
\sigma_{i_t}(AB) \ge \sum_{t=1}^k \sigma_{i_t}(A)
\sigma_{n-t+1}(B).\end{equation}
\end{lemma}
Set $n=4$ in both inequalities. Then put $k=1$ in the first, and
$k=3,i_1=2,i_2=3,i_3=4$ in the second. Subtracting the
inequalities then gives: \begin{eqnarray}
\sigma_1(AB)-(\sigma_2(AB)+\sigma_3(AB)+\sigma_4(AB))\leq\hspace{1.8cm} \nonumber\\
\sigma_1(A)\sigma_1(B)
-\sigma_2(A)\sigma_4(B)-\sigma_3(A)\sigma_3(B)-\sigma_4(A)\sigma_2(B)\nonumber\end{eqnarray}
Furthermore, let $A=\Lambda^{1/2}$ and $B=V\Lambda^{1/2}$, with
$\Lambda$ positive diagonal and with the diagonal elements sorted
in non-ascending order. Thus,
$\sigma_i(A)=\sigma_i(B)=\sqrt{\lambda_i}$. This gives:
$$
(\sigma_1-(\sigma_2+\sigma_3+\sigma_4))(\Lambda^{1/2} V
\Lambda^{1/2}) \le \lambda_1 -
(2\sqrt{\lambda_2\lambda_4}+\lambda_3).
$$
It is easy to see that this inequality becomes an equality iff
$V$ is equal to the permutation matrix
\begin{equation}\left(\begin{array}{cccc}
1 & 0 & 0 & 0 \\
0 & 0 & 0 & 1 \\
0 & 0 & 1 & 0 \\
0 & 1 & 0 & 0 \end{array}\right)\end{equation} multiplied by an
arbitrary unitary diagonal matrix $D_\phi$. Therefore, we have
proven:
\begin{eqnarray}
\nonumber &\max_{V\in U(4)}&(\sigma_1-(\sigma_2+\sigma_3+\sigma_4))(\Lambda^{1/2} V \Lambda^{1/2}) = \\
&&\hspace{2cm}\lambda_1 - (2\sqrt{\lambda_2
\lambda_4}+\lambda_3).\end{eqnarray} We can directly apply this to
the problem at hand. The optimal $V$ is indeed symmetric, so that
it can be decomposed as $V=V_1^T V_1$. A possible Takagi
factor is: \begin{equation} V_1 = \left(\begin{array}{cccc} 1 & 0 & 0 & 0 \\
0 & 1/\sqrt{2} & 0 & 1/\sqrt{2} \\
0 & 0 & 1 & 0 \\
0 & i/\sqrt{2} & 0 & -i/\sqrt{2}\end{array}\right)\end{equation}
The optimal unitary operations $U$ are thus all of the form: $U =
S_1^\dagger O V_1D_\phi^{1/2} \Phi^\dagger$ with $O$ an arbitrary
orthogonal matrix. It has to be emphasized that the diagonal
matrix $D_\phi$ will not have any effect on the state
$\rho'=U\Phi\Lambda\Phi^\dagger U^\dagger$.

To proceed we exploit a well-known accident in Lie group theory
:\begin{equation}SU(2)\otimes SU(2)\cong SO(4).\end{equation} It
now happens that the unitary matrix $S_1$ is exactly of the form
for making $S_1 (U_1\otimes U_2)S_1^\dagger$ real for arbitrary
$\{U_1,U_2\} \in SU(2)$. It follows that $S_1(U_1\otimes
U_2)S_1^\dagger$ is orthogonal and thus is an element of $SO(4)$.
Conversely, each element $Q\in SO(4)$ can be written as
$Q=S_1(U_1\otimes U_2) S_1^\dagger$. On the other hand the
orthogonal matrices with determinant equal to $-1$ can all be
written as an orthogonal matrix with determinant $1$ multiplied
by a fixed matrix of determinant $-1$. Some calculations reveal
that
\[S_1^\dagger
\left(\begin{array}{cccc}1&0&0&0\\0&1&0&0\\0&0&1&0\\0&0&0&-1\end{array}\right)V_1=(\sigma_y\otimes\sigma_y)S_1^\dagger
V_1\left(\begin{array}{cccc}1&0&0&0\\0&1&0&0\\0&0&-1&0\\0&0&0&1\end{array}\right)\]
We conclude that for each $O\in O(4)$ and $D_\phi$ unitary
diagonal, there exist $U_1,U_2\in SU(2)$ and $D_{\phi'}$ unitary
diagonal, such that $U=S_1^\dagger
OV_1D_\phi\Phi^\dagger=(U_1\otimes U_2)S_1^\dagger V_1
D_{\phi'}\Phi^\dagger$.

It is now easy to check that a unitary transformation produces
maximal entanglement of formation if and only if it is of the
form \begin{equation}\left(U_1\otimes
U_2\right)\left(\begin{array}{cccc}0&0&0&1\\1/\sqrt{2}&0&1/\sqrt{2}&0\\1/\sqrt{2}&0&-1/\sqrt{2}&0\\0&1&0&0\end{array}\right)D_\phi\Phi^{\dagger}.\end{equation}
This completes the proof of the first part of the Theorem.

We now proceed to prove the second part of the Theorem concerning
the negativity. This proof is based on the Rayleigh-Ritz
variational characterization of the minimal eigenvalue of a
Hermitian matrix:\begin{eqnarray}\lambda_{min}(\rho^{T_A})
&=& \min_{x: ||x||=1} \trace \rho^{T_A} \ket{x}\bra{x} \nonumber\\
&=& \min_{x: ||x||=1} \trace \rho
(\ket{x}\bra{x})^{T_A}\end{eqnarray} The eigenvalue decomposition
of $(\ket{x}\bra{x})^{T_A}$ can best be deduced from its singular
value decomposition. Let $\tilde{x}$ denote a reshaping of the
vector $x$ to a $2\times 2$ matrix with $\tilde{x}_{ij}=\langle
e^i\otimes e^j|x\rangle$. Introducing the permutation matrix
$P_0=\sum_{ij}e^{ij}\otimes e^{ji}$, the partial transpose can be
written as follows: \begin{equation} (\ket{x}\bra{x})^{T_A} = P_0
(\tilde{x} \otimes \tilde{x}^\dagger).\end{equation} The proof of
this statement is elementary. We denote the Schmidt decomposition
of the vector $\ket{x}$ by \begin{equation} \tilde{x} = U_1 \Sigma
U_2^\dagger,\end{equation} where the diagonal elements of
$\Sigma$ are given by $\sigma_1,\sigma_2$. Since $x$ is
normalized we can parameterize these as
$\cos(\alpha),\sin(\alpha)$ with $0\leq\alpha\leq\pi/4$ (to
maintain the ordering). We get
\begin{equation}\label{eq:svdx} (\ket{x}\bra{x})^{T_A} = P_0
(U_1\otimes U_2)(\Sigma\otimes\Sigma)(U_2\otimes
U_1)^\dagger.\end{equation} This clearly is a singular value
decomposition. The explicit eigenvalue decomposition can now be
calculated using the basic property of $P_0$ that $P_0(A\otimes
B)=(B\otimes A)P_0$ for arbitrary $A,B$. It is then easy to check
that the eigenvalue decomposition of $(\ket{x}\bra{x})^{T_A}$ is
given by:
\begin{equation}
(\ket{x}\bra{x})^{T_A}=V(x)D(\alpha(x))V(x)^{\dagger}
    \end{equation}
where $D(\alpha(x))$ is the diagonal matrix with eigenvalues
$(\sigma_1^2,\sigma_1\sigma_2,\sigma_2^2,-\sigma_1\sigma_2)$ and
\begin{equation}
V(x)=(U_1(x)\otimes U_2(x))\left(
    \begin{array}{cccc}
        1 & 0 & 0 & 0 \\
        0 & 1/\sqrt{2} & 0 & 1/\sqrt{2} \\
        0 & 1/\sqrt{2} & 0 & -1/\sqrt{2} \\
        0 & 0 & 1 & 0
    \end{array}
    \right)
\end{equation}
For the problem at hand, we have to minimize the minimal
eigenvalue of $(U\rho U^\dagger)^{T_A}$ over all possible $U\in
U(4)$. Thus, we have to minimize: \begin{eqnarray}
\min_{U,x} \trace U\Phi\Lambda \Phi^\dagger U^\dagger &V(x)& D(\alpha(x))V(x)^\dagger \nonumber\\
&=& \min_\alpha \min_W \trace \Lambda W^\dagger D(\alpha)
W,\end{eqnarray} where we have absorbed the eigenvector matrix
$\Phi$ of $\rho$, as well as $V(x)^\dagger$, into $U$, yielding
$W$. Now, the minimization over $W$ can be done by writing the
trace in components \begin{eqnarray}
g(\alpha) &=& \trace \Lambda W^\dagger D(\alpha) W \nonumber\\
&=& \sum_{i,j} d_j(\alpha) |W_{ji}|^2 \lambda_i \nonumber\\
&=& d(\alpha)^T J(W) \lambda,\end{eqnarray} where $d(\alpha)$ and
$\lambda$ denote the vectors containing the diagonal elements of
$D(\alpha)$ and $\Lambda$, respectively. $J(W)$ is a doubly
stochastic matrix formed from $W$ by taking the modulus squared
of every element. The minimum over all $W$ is attained when $J(W)$
is a permutation matrix; this follows from Birkhoff's theorem
\cite{horn85}, which says that the set of doubly-stochastic
matrices is the convex closure of the set of permutation matrices,
and also of the fact that our object function is linear. Since the
components of $\sigma$ and $\lambda$ are sorted in descending
order and $\lambda$ is positive, the permutation matrix yielding
the minimum for any $\alpha$ is the matrix
\begin{equation}\label{eq:J0} J_0=\left(\begin{array}{cccc}
0&0&0&1 \\
0&0&1&0 \\
0&1&0&0 \\
1&0&0&0
\end{array}\right).\end{equation}
Thus $W$ has to be chosen equal to $J_0$ multiplied by a diagonal
unitary matrix $D_\phi$. Hence, the minimum over $W$ is given by
$\sum_{j=1}^4 \lambda_j d_{4+1-j}(\alpha)$. Minimizing over
$\alpha$ gives, after a few basic calculations: \begin{eqnarray}
\cos(2\alpha) &=& \frac{\lambda_2-\lambda_4}{\sqrt{(\lambda_1-\lambda_3)^2+(\lambda_2-\lambda_4)^2}}\nonumber\\
g(\alpha)
&=&\left(\lambda_2+\lambda_4-\sqrt{(\lambda_1-\lambda_3)^2+(\lambda_2-\lambda_4)^2}\right)/2.\nonumber\end{eqnarray}
This immediately yields the conjectured formula for the optimal
negativity.

We now have to find the $U$ for which this optimum is reached. As
$V(x)^\dagger U\Phi=W$, it follows that the optimal unitary
transformation $U$ is given by $U=V(x)J_0D_\phi\Phi^\dagger$:
\begin{equation} U=\left(U_1\otimes
U_2\right)\left(\begin{array}{cccc}0&0&0&1\\1/\sqrt{2}&0&1/\sqrt{2}&0\\1/\sqrt{2}&0&-1/\sqrt{2}&0\\0&1&0&0\end{array}\right)D_\phi\Phi^{\dagger}\end{equation}
This is exactly the same $U$ as in the case of entanglement of
formation.

Next we move to the third part of the theorem concerning the
relative entropy of entanglement. We first prove two lemmas.

\begin{lemma} Consider the class of superoperators
$$
{\cal T}(\rho) = \sum_i a_i U_i \rho U_i^\dagger,
$$
where all $U_i$ are unitary, and the $a_i$ form a distribution.
Then, for any state $\rho$ that is invariant under $\cal T$, we
have for the relative entropy:
$$
S(\rho||\sigma) \ge S(\rho||{\cal T}^\dagger(\sigma)).
$$
\end{lemma}
{\bf Proof.} The proof of this lemma is heavily inspired by
theorem 6 in Rains \cite{Rains}. From $S(\rho)=S({\cal T}(\rho))$,
we find
\begin{eqnarray*}
S(\rho||\sigma) &=& \trace\rho\log\rho-\trace\rho\log\sigma \\
&=&\trace\rho\log\rho-\trace{\cal T}(\rho)\log\sigma \\
&=&\trace\rho\log\rho-\sum_i a_i\trace U_i\rho U_i^\dagger \log\sigma \\
&=&\trace\rho\log\rho-\sum_i a_i\trace \rho \log(U_i^\dagger\sigma U-i) \\
&\ge& \trace\rho\log\rho-\trace \rho \log(\sum_i a_i U_i^\dagger\sigma U-i) \\
&=& S(\rho||{\cal T}^\dagger(\sigma)),
\end{eqnarray*}
where in the penultimate line, we have used the subadditivity of
the relative entropy w.r.t.\ its second argument.

\begin{lemma} For $\rho$ of the form $\rho=U\Lambda U^\dagger$
with \[
U=\left(\begin{array}{cccc}0&0&0&1\\1/\sqrt{2}&0&1/\sqrt{2}&0\\1/\sqrt{2}&0&-1/\sqrt{2}&0\\0&1&0&0\end{array}\right)
\] and $\Lambda$ containing the ordered eigenvalues of $\rho$,
$$
E_R(\rho) = \min_{\sigma\in D\cap\mbox{MEMS}} S(\rho||\sigma).
$$ where MEMS is the class of maximally entangled mixed states.
\end{lemma}
{\bf Proof.} Define the following superoperator:
$$
{\cal T}(\rho) = V \diag(V^\dagger\rho V) V^\dagger.
$$
Here, $\diag(\rho)$ is the superoperator that sets all
off-diagonal elements of $\rho$ equal to zero while keeping the
diagonal ones intact. This superoperator can also be written as
$$
\diag(\rho) = \sum_i P_i \rho P_i/2^n,
$$
where $P_i$ runs through all possible diagonal matrices having
only $+1$ or $-1$ on their diagonal \cite{olkin}. It follows that
${\cal T}$ is of the form mentioned in the first Lemma and,
furthermore, that it is a self-dual superoperator, i.e.\ ${\cal
T}^\dagger={\cal T}$.

It is obvious that all MEMS states (with $U_1=U_2=\identity$) are
left invariant by $\cal T$. We will now show that any such $\cal
T$ maps separable states to separable states. Consider thereto
the pure product states only; if the proposition is valid for
pure product states, it will be valid for all separable states
(by linearity). The most general pure product state has the state
vector $\psi=(ac,ad,bc,bd)$, with $a,b,c,d$ complex numbers.
Then, since
$$
V^\dagger\psi = ((ad+bc)/\sqrt{2},bd,(ad-bc)/\sqrt{2},ac),
$$
\begin{eqnarray*}
{\cal T}(\psi\psi^\dagger) &=& V \diag(V^\dagger \psi\psi^\dagger V)V^\dagger \\
&=& V \Lambda V^\dagger,
\end{eqnarray*}
where the diagonal elements of $\Lambda$ are, in order,
$$
(|(ad+bc)/\sqrt{2}|^2,|bd|^2,|(ad-bc)/\sqrt{2}|^2,|ac|^2).
$$
As these values are not necessarily sorted, ${\cal
T}(\psi\psi^\dagger)$ need not be MEMS. However, it is still
possible to apply the formula for the negativity of MEMS states
which says that
$$
E_N(\rho) =
\max(0,\sqrt{(\lambda_1-\lambda_3)^2+(\lambda_2-\lambda_4)^2}-\lambda_2-\lambda_4).
$$
As can be easily checked, the validity of this formula does not
rely on the ordering of the $\lambda_i$, as long as each
$\lambda_i$ pertains to the $i$-th column of $V$. In particular,
using
\begin{eqnarray*}
\lambda_1-\lambda_3 &=& 2\Re(ab(cd)^*) \\
\lambda_2\pm\lambda_4 &=& |bd|^2\pm|ac|^2,
\end{eqnarray*}
we get for the negativity of ${\cal T}(\psi\psi^\dagger)$:
$$
E_N({\cal T}(\psi\psi^\dagger)) = \max(0,F)
$$
with
\begin{eqnarray*}
F&=& \sqrt{(\lambda_1-\lambda_3)^2+(\lambda_2-\lambda_4)^2}-(\lambda_2+\lambda_4) \\
&=& \sqrt{4\Re(ab(cd)^*)^2+|bd|^4+|ac|^4-2|abcd|^2}\\
&&  -(|bd|^2+|ac|^2) \\
&=& \sqrt{|bd|^4+|ac|^4+2|abcd|^2 - 4\Im(ab(cd)^*)^2}\\
&&  -(|bd|^2+|ac|^2) \\
&=& \sqrt{(|bd|^2+|ac|^2)^2 - 4\Im(ab(cd)^*)^2}-(|bd|^2+|ac|^2) \\
&\le& 0.
\end{eqnarray*}
Hence, ${\cal T}(\psi\psi^\dagger)$ is separable, as we set out
to prove, so that $\cal T$ maps separable states to separable
states.

From the previous discussion it also follows that states of the
form $V\Lambda V^\dagger$ are separable if and only if the
eigenvalues satisfy
\begin{equation} \label{eqsep}
\sqrt{(\lambda_1-\lambda_3)^2+(\lambda_2-\lambda_4)^2}-\lambda_2-\lambda_4
\le 0
\end{equation}
Furthermore, states $V\Lambda V^\dagger$ are obviously invariant
under $\cal T$. Hence, letting $\sigma$ traverse all separable
states of this form generates the same set of states ${\cal
T}(\sigma)$ as letting $\sigma$ traverse all separable states
without restriction. Therefore,
\begin{eqnarray*}
E_R(\rho) &=& \min_{\sigma\in D} S(\rho||\sigma) \\
&\ge& \min_{\sigma\in D} S(\rho||{\cal T}(\sigma)) \\
&=& \min_{\sigma=V\Lambda V^\dagger \in D} S(\rho||\sigma).
\end{eqnarray*}
Comparing the first and the third line, we immediately see that
the inequality must be an equality.

Actually, an even stronger result holds, as we can restrict
ourselves in this minimization to states $\sigma=V\Lambda
V^\dagger \in D$ where the diagonal elements appear in descending
order ($\lambda_1\ge\lambda_2\ge\lambda_3\ge\lambda_4$). In other
words, $\sigma$ may be taken from the set of separable MEMS
states. To see this, note that, as $\rho$ and $\sigma$ are both
MEMS,
$$
S(\rho||\sigma) = \sum_i p_i (\lg p_i - \lg \lambda_i),
$$
where the $p_i$ are the sorted eigenvalues of $\rho$, and
$\lambda_i$ are the not necessarily sorted eigenvalues of
$\sigma$. It is easy to see that one always gets a lower relative
entropy by permuting the $\lambda_i$ into descending order
\cite{olkin}. This ends the proof of the lemma.

It is now easy to prove the last part of the main theorem: Because
the $\sigma$ are restricted to separable MEMS states, this means
that, for any global unitary $U$, $U\sigma U^\dagger$ is still
separable. Hence, for $\rho \in {\rm MEMS}$,
\begin{eqnarray*}
E_R(\rho) &=& \min_{\sigma\in D\cap\mbox{MEMS}} S(\rho||\sigma) \\
&=& \min_{\sigma\in D\cap\mbox{MEMS}} S(U\rho U^\dagger|| U\sigma
U^\dagger) \\
&\ge & \min_{U\sigma U^\dagger\in D} S(U\rho U^\dagger|| U\sigma
U^\dagger) \\
&=& E_R(U\rho U^\dagger),
\end{eqnarray*}
where the inequality in the penultimate line arises because the
minimization domain has been enlarged. Therefore the MEMS states
have larger relative entropy of entanglement then all states that
can be obtained from it by doing global unitary operations.

The explicit calculation of the relative entropy of entanglement
of the maximally entangled mixed states is now a tedious but
straightforward exercise, whose result is quoted in the theorem.
This completes the proof of Theorem 1. \qed

Let us now analyze more closely the newly defined class of
generalized Bell states. We already know that $U$ is unique up to
local unitary transformations. It is easy to check that the
ordered eigenvalues of the generalized Bell states for given
entanglement of formation $f(C)$ are parameterized by two
independent
variables $\alpha$ and $\beta$: \begin{eqnarray}0&\leq&\alpha\leq 1\nonumber\\
\beta&\geq&\sqrt{1-\frac{\alpha^2}{9}}-\sqrt{\frac{8}{9}}\alpha\nonumber\\
\beta&\leq&\min(\sqrt{\frac{1+C}{1-C}-\frac{\alpha^2}{9}}-\sqrt{\frac{2}{9}}\alpha,
\sqrt{3-\alpha^2}-\sqrt{2}\alpha)\nonumber\\
\lambda_1&=&1-\frac{1-C}{6}(3+\beta^2)\nonumber\\
\lambda_2&=&\frac{1-C}{6}(\alpha+\sqrt{2}\beta)^2\nonumber\\
\lambda_3&=&\frac{1-C}{6}(3-(\sqrt{2}\alpha+\beta)^2)\nonumber\\
\lambda_4&=&\frac{1-C}{6}\alpha^2\end{eqnarray} For given EoF
there is thus, up to local unitary transformations, a two
dimensional manifold of maximally entangled states. In the case of
concurrence $C=1$ the upper and lower bounds on $\beta$ become
equal and the unique pure Bell states arise. Another observation
is the fact that $\lambda_4$ of all generalized Bell states is
smaller then $1/6$. This implies that if the smallest eigenvalue
of whatever two-qubit state exceeds $1/6$, this state is
separable.

A natural question is now how to characterize the entangled states
closest to the maximally mixed state. A sensible metric is given
by the Frobenius norm
$\|\rho-\identity\|_2=\sqrt{\sum_i\lambda_i^2-1/4}$. This norm is
only dependent on the eigenvalues of $\rho$ and it is thus
sufficient to consider the generalized Bell states at the
boundary of entangled states where both the concurrence and the
negativity become zero. This can be solved using the method of
Lagrange multipliers. A straightforward calculation leads to a
one-parameter family of solutions:
\begin{eqnarray}
0\leq&x\leq&\frac{1}{6}\nonumber\\
\lambda_1=\frac{1}{3}+\sqrt{x\left(\frac{1}{3}-x\right)}&&
\lambda_2=\frac{1}{3}-x\nonumber\\
\lambda_3=\frac{1}{3}-\sqrt{x\left(\frac{1}{3}-x\right)}
&&\lambda_4=x
\end{eqnarray}
The Frobenius norm $\|\rho-\identity\|_2$ for all these states on
the boundary of the sphere of separable states is given by the
number $\sqrt{1/12}$. This criterion is exactly equivalent to the
well-known criterion of Zyczkowski et al. \cite{zyczkowski}:
$\trace{\rho^2}=1/3$. Here, however, we have the additional
benefit of knowing exactly all the entangled states on this
boundary as these are the generalized Bell states with
eigenvalues given by the previous formula. Furthermore,
Zyczkowski et al. \cite{zyczkowski} proposed a lower bound on the
volume of separable states by considering the ball of states that
remain separable under all global unitary transformations.
Clearly the criterion $\sum_i\lambda_i^2\leq 1/3$ can be
strengthened to
$\lambda_1-\lambda_3-2\sqrt{\lambda_2\lambda_4}\leq 0$. Some
tedious integration then leads to a better lower bound for the
volume of separable states relative to the volume of all states:
$0.3270$ (as opposed to $0.3023$ of \cite{zyczkowski}).

Further interesting properties of the maximally entangled mixed
states include the fact that the states with maximal entropy for
given entanglement all belong to this class. This can be seen as
follows: the global entropy of a state is a function of the
eigenvalues of the density matrix only. Therefore the states with
maximal entanglement for given entropy can be found by first
looking for the states with maximal entanglement for fixed
eigenvalues, followed by maximizing the entropy of the obtained
class of (maximally entangled) mixed states.

In conclusion, we have generalized the concept of pure Bell states
to mixed states of two qubits. We have proven that the
entanglement of formation, the negativity and the relative
entropy of entanglement of these generalized Bell states could
not be increased by applying any global unitary transformation.
Whether their entanglement of distillation is also maximal is an
interesting open problem.

We thank Lieven Vandersypen for bringing the problem under our
attention, and J. Dehaene, L. De Lathauwer and K. Zyczkowski for
valuable comments. T. De Bie is a Research Assistant with the Fund
for Scientific Research-Flanders (FWO-Vlaanderen).

\end{multicols}
\end{document}